\documentstyle[11pt,newpasp,twoside,epsf,psfig]{article}
\def\edcomment#1{\iffalse\marginpar{\raggedright\sl#1\/}\else\relax\fi}
\reversemarginpar
 
\begin{document}
\title{The Universe behind the Southern Milky Way}
 \author{Ren\'ee C. Kraan-Korteweg}
\affil{Depto.\ de Astronom\' \i a, Universidad de Guanajuato,
Apdo.~Postal 144, Guanajuato, GTO 36000, Mexico}
\author{Lister Staveley-Smith, Jennifer Donley}
\affil{Australia Telescope National Facility, CSIRO, P.O. Box 76, Epping,
NSW 1710, Australia}
\author{Patricia A. Henning}
\affil{Institute for Astrophysics, University of New Mexico,
800 Yale Blvd., NE, Albuquerque, NM 87131, USA}

\begin{abstract} A first analysis of a deep blind HI 
survey covering the southern Zone of Avoidance plus an extension
towards the north ($196\deg \le \ell \le 52\deg$) obtained with the
Multibeam receiver at the 64\,m Parkes telescope reveals slightly over
a thousand galaxies within the latitude completeness limit of $|b| \le
5\deg$. The characteristics and the uncovered large-scale structures
of this survey are described, in particular the prominence of the
Norma Supercluster, the possible cluster around PKS\,1343$-$601 (both
in the Great Attractor region), as well as the Local Void and the 
clustering in the Puppis region.

In this blind HI survey, HIZOA\,J0836$-$43, one of the most massive
spiral galaxies known to date was discovered ($M_{\rm HI} = 7.3 \cdot
10^{10}{\rm M_\odot}$, $M_{\rm T} = 1.1 \cdot 10^{12}{\rm M}_\odot$;
H$_0 = 75$\,km/s/Mpc). Although of similar mass as Malin 1-like objects,
this galaxy does not share their typical low-surface brightness
properties but seems an exceptionally massive but normal, high-surface
brightness, star-forming galaxy.
\end{abstract}

\section{Introduction}
Based on the intensive multi-wavelength surveys of the Zone of
Avoidance (ZOA) in the last decade -- aimed at addressing cosmological
questions about the dynamics of the Local Group, respectively the
possible existence of nearby hidden massive galaxies, dipole
determinations based on luminous galaxies, the continuity and size of
nearby superclusters, the mapping of cosmic flow fields -- enormous
progress has been achieved in uncovering the galaxy distribution in
this obscured region of the sky (see Kraan-Korteweg \& Lahav 2000 for
a review).

In particular, the near infrared, whole-sky homogeneous surveys such
as 2MASS (see Huchra et al., these proceedings) have resulted in a
magnificent reduction of the ZOA. However, even this survey does
become increasingly incomplete at low Galactic latitudes in the larger
Galactic Bulge (GB) area ($\ell \approx 0\deg \pm 90\deg$), including
the Great Attractor (GA) region.  Even if obscured galaxies can be
identified, redshifts are difficult if not impossible to obtain at the
higher extinction levels.

Because of the transparency of the Galaxy to the 21\,cm radiation of
neutral hydrogen, systematic HI-surveys are particularly powerful in
mapping large-scale structure (LSS) in this part of the sky. The
redshifted 21\,cm emission of HI-rich galaxies are readily detectable
at lowest latitudes and highest extinction levels and the signal will
furthermore provide immediate redshift and rotational velocity
information.

\section{The Parkes Multibeam HI ZOA Survey}
For these reasons, a systematic deep blind HI survey of the southern
Milky Way was begun in 1997 with the Multibeam receiver at the 64\,m
Parkes telescope. The data presented here consist of the preliminary
analysis of 27 consecutive slightly overlapping data cubes, offset by
$\Delta \ell = 8\deg$, centered on the southern Galactic Plane (GP):
$196\deg \le \ell \le 52\deg$, $|b| \le 5\deg$. The 23 central cubes
cover the {\em southern} Milky Way (Henning et al., in prep.), while
the other four are extensions to the north (Donley et al., in
prep.). The coverage in redshift space is $-1200 \la v_{\rm hel} \la
12700$\,km/s. With an integration time of 25~min/beam -- respectively 25
scans -- an rms = 6mJy is achieved, making this survey sensitive to
the lowest mass dwarf galaxies in the local neighborhood and to
normal spirals well beyond the GA region.

Results based on a shallow subsample of this survey (2 out of 25
scans) were presented in Henning et al. 2000 (henceforth HIZSS)
and resulted in the detection of 110 mainly nearby galaxies, 67 of
which were previously unknown.

\section{The Detected Galaxies}
A careful analysis of the 27 data cubes by at least two, sometimes
three, individual researchers with the subsequent neutral evaluation
of inconsistent cases in the detection lists by a third party yields
1075 galaxies. The final list may vary slightly. Thirty-nine of these
galaxies have latitudes above $|b| = 5\deg$. As the sensitivity
decreases for $|b| > 5\deg$, and therewith the completeness, these 39
galaxies are not included in the presented graphs and discussions.

\begin{figure}[t] 
\plotone{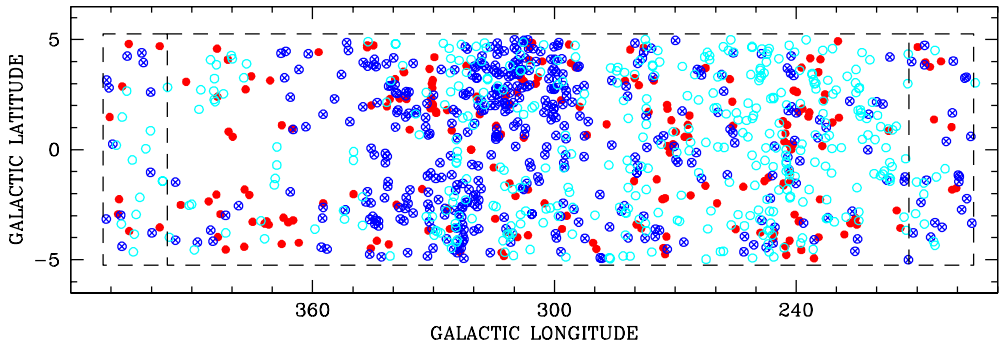}
\caption{Distribution in Galactic coordinates of the 1036 galaxies
detected in the deep HI ZOA survey. Open circles: $v_{\rm hel} <
3500$; circled crosses: $3500 < v_{\rm hel} < 6500$; filled circles:
$v_{\rm hel} > 9500$\,km/s.}
\label{sky}
\end{figure}

Figure~1 displays the distribution along the Milky Way of the in HI
detected galaxies. An inspection of this distribution shows that the
HI survey nearly fully penetrates the ZOA with hardly any dependence
on Galactic latitude. This is confirmed by the left panel in Fig.~2,
which shows the detection rate as a function of Galactic latitude. The
small dip in the detection rate between $-2\deg \la b \la +1\deg$
stems mainly from the GB regions and to a lesser extent from the GA
region ($\ell \approx 300\deg-340\deg$). The former is due to the high number
of continuum sources at low latitudes in the GB region, while the latter
probably is due to the high galaxian density where a moderate number
of continuum sources may already result in a detection-loss for the on
average higher velocity, hence fainter, GA galaxies. This explanation
is supported by the fact that this dip is not noticeable in the
shallower HIZSS data (lower histogram).

A much stronger variation is apparent in the number density as a
function of Galactic longitude (see also middle panel of
Fig.~2). This can be explained entirely with large-scale structure
(see also next section) such as the nearby -- and therefore in HIZSS
prominent -- Puppis filament ($\ell \approx 240\deg$), the
Hydra-Antlia filament ($\ell \approx 280\deg$), the very dense GA
region ($\ell \approx 300-340\deg$), followed by an underdense region
($52\deg \ga \ell \ga 350\deg$) which is strongly influenced by the Local
and Sagittarius Void (see also HIZSS histogram). 

These LSS have left their imprint also on the velocity diagram (right
panel of Fig.~2), which shows two conspicuous broad peaks. The
low-velocity one is due to a blend of various structures in the GP or
crossing the GP (see next section) while the second around 5000\,km/s
clearly is due to the GA overdensity, respectively the Norma
Supercluster (see also Fig.~3).  The velocity histogram moreover shows
that galaxies are found all the way out to the velocity limit of
the survey of $\sim 12000$\,km/s, hence probe the galaxy distribution
considerably deeper than HIZSS, the HI Bright Galaxy Catalog BGC
(Koribalski et al.\ 2003) or the HI Parkes All Sky Survey HIPASS (see
Zwaan et al., these proceedings).

\begin{figure}[h]
\plotone{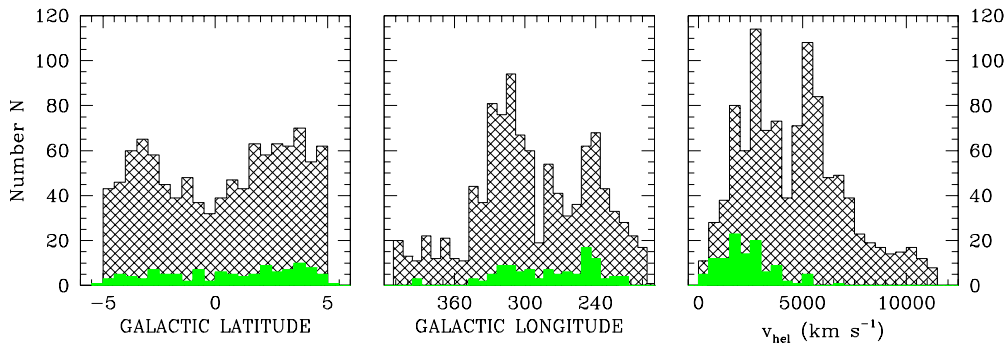}
\caption{Distributions as a function of the Galactic latitude, longitude 
and the heliocentric velocity of the 1036 in HI detected galaxies. 
The lower histograms represent the results from the HIZSS.}
\label{hist}
\end{figure}

\section{Uncovered Large-Scale Structures}

Figure 3 shows a Galactic latitude slice with $|b| \le 5\deg$ out
to 12000\,km/s of the 1035 galaxies detected in the deep Parkes HIZOA 
survey within $196\deg \le \ell \le 52\deg$. The first very
obvious conclusion when inspecting this figure is that, yes, the HI
survey really permits to trace LSS in the most opaque part of the ZOA
in a homogenous way, unbiased by the clumpiness of the foreground dust
contamination and shows good coverage out to about 7000\,km/s.

\begin{figure}[t] 
\hfil \psfig{figure=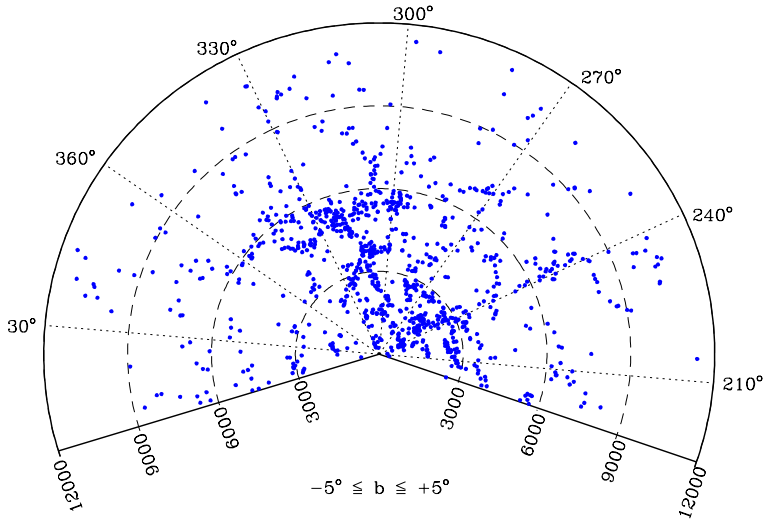,width=10cm} \hfil
\caption{Galactic latitude slice with $|b| \le 5\deg$ out to
12000\,km/s of the 1035 in HI detected galaxies. Circles mark
intervals of 3000\,km/s.}
\label{wedge}
\end{figure}

While discussing some of the most interesting features revealed with
Fig.~3, it is suggested to simultaneously consult Fig.~4, which shows
the HIZOA data with data extracted from LEDA surrounding the ZOA in
sky projections for three velocity shells of thickness 3000\,km/s to
show the newly discovered features in context to known structures.

\begin{figure}[t]
\hfil \psfig{figure=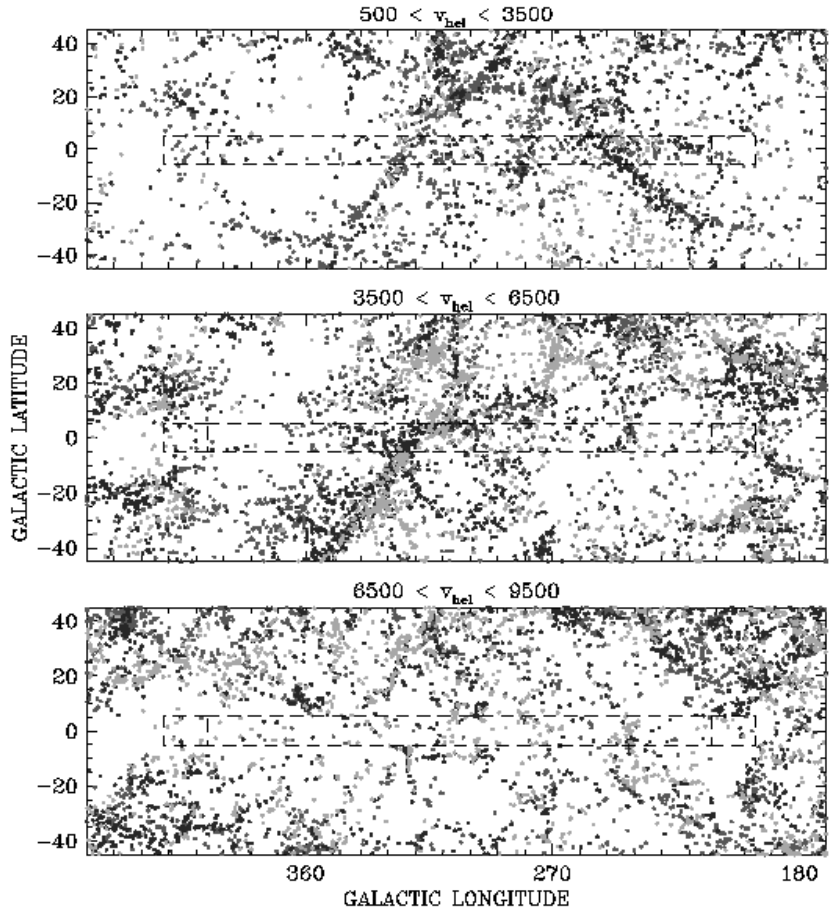,width=11cm} \hfil
\caption{Sky projections of three redshift slices of depth 
$\Delta v = 3000$\,km/s showing the HIZOA data in combination with 
data from LEDA. The HIZOA survey area is outlined.}
\label{vslice}
\end{figure}

The most prominent LSS in Fig.~3 certainly is the {\em Norma Supercluster}
which seems in this plot to stretch from $360\deg$ to $290\deg$, lying
always just below the 6000\,km/s circle, with a weakly visible
extension towards Vela ($\sim 270\deg$) at 6-7000\,km/s. The latter
is stronger at latitudes above the HIZOA (see panel 2 and 3 in
Fig.~4). This wall-like feature seems to consist of various
concentrations, where the one around $340\deg$ is seen for the first
time in the HIZOA. Because of the high extinction there, it cannot be
assessed whether this overdensity extends further than $|b| =
5\deg$. The clump at $325\deg$ is due to the low-latitude side of the
Norma cluster A3627 at ($\ell,b,v = 325\deg,-7\deg,4880$\,km/s;
Kraan-Korteweg et al.\,1996). The next two due to groups/clusters at
$310\deg$ and $300\deg$ respectively, both at $|b| \sim 4\deg$,
are very distinct in the middle panel of Fig.~4. Between these two
clusters at slightly higher latitude we see a small finger of God,
which belongs to the by Woudt (1998) identified Centaurus-Crux
cluster.

This is not the only overdensity in the GA region. A significant
agglomeration of galaxies is evident closer by at $(\ell,v) =
(311\deg,3900$\,km/s) with two filaments merging into it. Although no
finger of God is visible (never very noteable in HI redshift slices),
this concentration forms part of the by Kraan-Korteweg \& Woudt (1999)
suspected heavily obscured cluster ($b = 2\deg, A_{\rm B} = 12^{\rm
m}$) around the strong radio source PKS\,1343$-$601. A deep $I$-band
survey (Kraan-Korteweg et al., in prep.) as well $J, H, K$ imaging of
its core (Nagayama et al., in prep.) are consistent with this being an
intermediate size cluster similar to Hydra.

Although the overdensity in the GA region looks remarkable here, a
preliminary analysis of the 4 cubes covering $300\deg \le \ell \le
332\deg$ by Staveley-Smith et al.~2000 shows an excess mass above the
background of 'only' $\sim 2 \cdot 10^{15}\Omega_0 {\rm M_\odot}$.

To the left of the GA overdensity we find an underdense region
conformed of the {\em Local Void} and the {\em Sagittarius Void} at
about $\ell=360\deg$ and central velocities of $\sim 1500$ and
4500\,km/s (quite conspicous also in the first two panels of
Fig.~4). Except for the tiny group of galaxies at about
($350\deg,3000$\,km/s), the distribution here and in Fig.~4 rather
suggest one big void than two seperate ones.

The righthand side of Fig.~4 is quite crowded, in particular the
Puppis region ($\ell \sim 240\deg$) with its two nearby groups (800
and 1500\,km/s) followed by the Hydra Wall at about 3000\, km/s that
extends from the Monocerus group ($210\deg$) to the concentration at
$280\deg$. The latter is not the signature of a group but due to a
filament emerging out of the Antlia cluster ($273\deg,19\deg$; see
Fig.~4). Regarding the here recognized possible filamentary connection
between the Monocerus group and Antlia (top panel Fig.~4), the dinosaur
that left its footprint on our Universe must actually have been
four-toed and not three-toed as envisioned by Lynden-Bell.

\section{A Supermassive Spiral Galaxy}

While searching the data cube Z264, centered on $\ell=264\deg$, an
exceptionally strong signal ($S_{\rm peak} = 47$\,mJy) for its very
high velocity ($v_{\rm hel} =10700$\,km/s) was identified. With the
very broad width of the signal ($\Delta v \approx 600$\,km/s), the
data was indicative of a galaxy of HI mass $M_{\rm HI} = 6 \cdot
10^{10}{\rm M_\odot}$ (assuming $D=v_{\rm CMB}/{\rm H}_0$; H$_0
=75$\,km/s/Mpc). This would place it amongst the most massive spiral
galaxies known to date, in the range of, e.g., Malin~1 ($5 - 10
\cdot10^{10}{\rm M_\odot}$; Bothun et al.\ 1987; Pickering \& Impey
1997). Surprising, considering the fact that no equivalently massive
galaxy was uncovered in the recent systematic HI sky-survey BGC (N =
1000 galaxies) and only one with $M_{\rm HI} > 5 \cdot 10^{10}{\rm
M_\odot}$ in HIPASS (N = 4300, Kilborn, priv. comm.), even though the
probability of detecting galaxies in this mass range is smaller for
the ZOA survey compared to both the BGC and HIPASS. Indeed the steep
exponential fall-off at the high mass end of the HI mass function
(HIMF) below $M_*$ (Zwaan et al.  2003, based on BGC; $M_* =
6\cdot10^{10}{\rm M_\odot}$, $\alpha = -1.30$, $\theta_* = 0.0086$,
see his Fig.~4) is so strong, that the above galaxy candidate would
lie well beyond the by them obtained HIMF -- though it must be
maintained that the volume densities of such high mass galaxies still
remain ill-constrained.

An optical counterpart could not be identified which actually was
hardly likely with a foreground extinction of $A_{\rm B} = 9\fm8$
(Schlegel et al.~1998) at the position in the sky of $(\ell,b) =
(262\fdg45,-1\fdg62)$. Both the DENIS and 2MASS infrared surveys
reveal, however, two possible weak counterparts.

\begin{figure}[h!!]
\hfil \psfig{figure=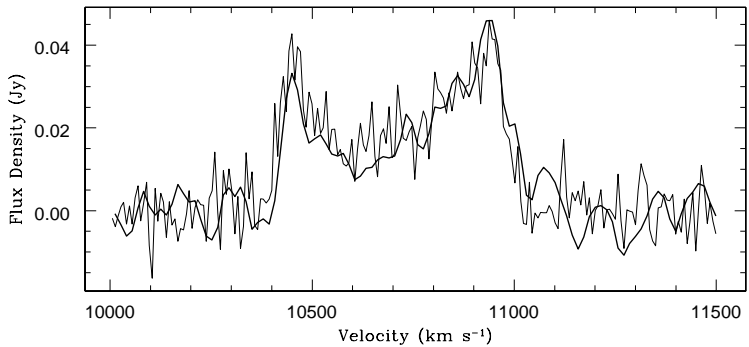,width=10cm} \hfil
\caption{HI profile of HIZOA\,J0836$-$43. Thick contour from Parkes 
HI ZOA survey; thin contour from ATCA follow-up observations.}
\end{figure}

The confirmation of this galaxy as an individual galaxy and therewith
its mass 
will help narrow down the number density of these rare
objects, an important value for galaxy formation and evolution models.

The galaxy candidate was then observed in February 2003 with the
Australia Telescope Compact Array ATCA (12\,hrs in 750D configuration;
6.6\,km/s velocity resolution after Hanning smoothing; Donley et al.,
in prep.). As can be verified from Fig.~5, which shows the profiles
from both observations, the ATCA spectral line profile is entirely
consistent with the Parkes MB detection. The galaxy HIZOA\,J0836$-$43
is in fact a single system (see Fig.~6) at a velocity of $v_{\rm hel}
= 10689$\,km/s. With a linewidth of $\Delta v_{\rm 20} = 610$\,km/s
and a flux of $I = 14.5$\,Jy\,km/s, the ATCA observations do confirm
the high mass of this object, providing a final value of $7.3 \cdot
10^{10}{\rm M_\odot}$.

In April 2003, the Anglo Australian Telescope (AAT) was used to obtain 
$K_s$- and $H$-band images. 
Figure~6 shows the 6 x 6 arcmin $K_s$-band image with the velocity
integrated HI-emission overlaid. There are two galaxies in the field
of view, both of which have counterparts in the 2MASS Extended Source
Catalog. The galaxy 2MASX\,J08365157$-$4337407, at an offset of only
$9\arcsec$ and a similar inclination angle as the HI gas, clearly is
the optical counterpart. The second visible galaxy seems at similar
distance but was not detected. This is not really surprising as this
galaxy, to have been detected with our ATCA observation, should have 
an HI-mass of over $M_{\rm HI} > 5 \cdot 10^{9}{\rm M_\odot}$ which is
high for a normal fairly face-on galaxy. The HI and $K_s$-band morphology 
do not show any obvious sign of interaction between the galaxies.

\begin{figure}[h!!]
\hfil \psfig{figure=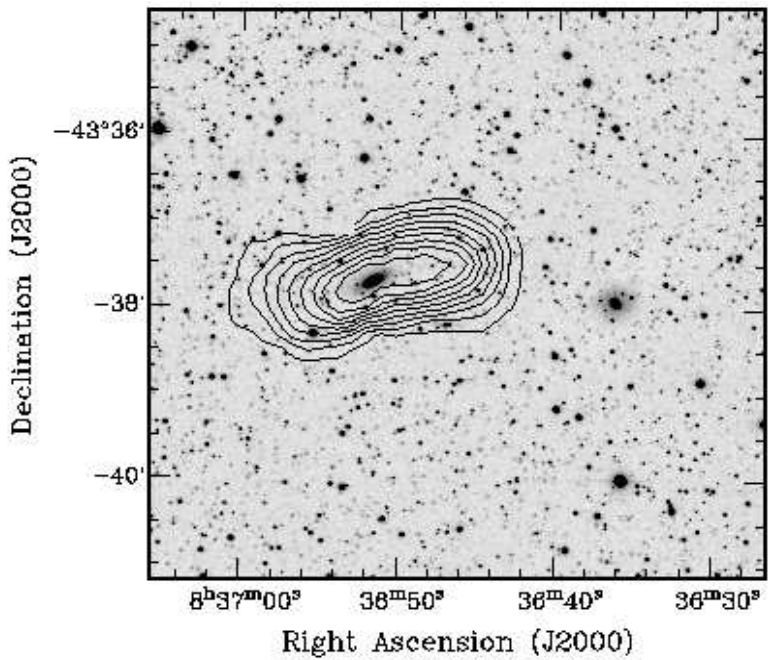,width=10.2cm,bbllx=94pt,bblly=279pt,bburx=452pt,bbury=589pt,clip=,angle=0} \hfil 
\caption{$K_s$-band image of HIZOA\,0836$-$43 from the AAT with the HI
contours obtained with ATCA superimposed. The HI contour level range
from 10\% to 90\% of the peak value. Synthesized beam: $58\arcsec
\times 41\arcsec$.}
\end{figure}

With an inclination of $i = 66\deg$ (from 
$K_s$-band photometry) and a linear radius of 50\,kpc,
the dynamical mass turns out to be $M_{\rm T} = 1.1 \cdot 10^{12}{\rm
M}_\odot$, also amongst the highest ever measured for spiral
galaxies. 

{\em Discussion:\ \ } Is this galaxy a new example of the giant LSB
class galaxies like Malin~1?  The observations clearly do not support
this. The rotation curve is steep in the center, remaining constant at
large radii, consistent with normal spiral galaxies having a dark
matter halo that extends well beyond the optical disk. The galaxy has
a HSB bulge of about 4\,kpc -- and a disk that can be traced out to
20\,kpc, even with the extinction in $K_s$ of $A_{\rm K} =
0\fm83$. Moreover, the HI density peak is a factor of $2 - 3$ higher
compared to values measured for LSB giant spiral galaxies by Pickering
et al.\,(1997) and the star formation rates determined from our radio
continuum data, as well as from IRAS far infrared fluxes (Cao et
al.\,2003) indicate a healthy star formation rate of about $20 - 30
{\rm M_\odot/yr}$.

HIZOA\,J0836$-$43 therewith has the properites of a normal HSB spiral
galaxy albeit of extreme mass. This is confirmed by the fact that it
fits perfectly on the $K$-band Tully--Fisher relation as determined by
Macri et al.\,(2003). As such, this galaxy is an interesting object to
study in further detail -- its intrinsic properties as well as its
environment -- in order to better understand how such a supermassive
galaxy could have formed by today within the current hierarchical
galaxy formation models.

\acknowledgments The contributions of the ZOA team (A.J. Green,
R.D. Ekers, R.F. Haynes, R.M. Price, E.M. Sadler, I. Stewart and
A. Schr\"{o}der) and other participants in this survey are gratefully
acknowledged. This research used the Lyon-Meudon Extragalactic
Database (LEDA), supplied by the LEDA team at the Centre de Recherche
Astronomique de Lyon, Obs.~de Lyon. RCKK thanks CONACyT for their
support (research grants 27602E and 40094F).

\end{document}